\documentclass[notitlepage,12pt,tightenlines,eqsecnum,nofootinbib,floatfix,prd,letterpaper,superscriptaddress]{revtex4-1}

\usepackage[utf8]{inputenc}
\usepackage{epsfig}
\usepackage{amsfonts}
\usepackage{amssymb}
\usepackage{amsbsy} 
\usepackage{amsmath}
\usepackage{amsthm}
\usepackage{graphicx}% Include figure files
\usepackage{dcolumn}% Align table columns on decimal point
\usepackage{bm}% bold math
\usepackage[nolist]{acronym}
\usepackage{graphics}
\usepackage[caption=false]{subfig}
\usepackage{lastpage}
\usepackage{fancyhdr}
\usepackage[english]{babel}
\usepackage{tabularx} 
\usepackage{mathptmx}
\usepackage[letterpaper, portrait, margin=1in]{geometry}
\usepackage{setspace}
\usepackage[breaklinks=true]{hyperref}% add hypertext capabilities
\usepackage{setspace}
\usepackage{color}

\usepackage{etoolbox}
\newtoggle{whitepaper}
\togglefalse{whitepaper}

\newtoggle{sections}
\toggletrue{sections}
\newacro{GW}{Gravitational wave}
\newacro{LIGO}{Laser Interferometer Gravitational-Wave Observatory}
\newacro{LISA}{\textit{Laser Interferometer Space Antenna}}
\newacro{MBH}{massive black hole}
\newacro{EMRI}{\emph{extreme mass-ratio inspiral}}
\newacro{EM}{electromagnetic}

\allowdisplaybreaks[4]

\AtBeginDocument{
  \addtocontents{toc}{\footnotesize}
  \addtocontents{lof}{\footnotesize}
}

\begin{document}

%%%%%%%%%%%%%%%%%%%%%%%%%%%%%%%%%%%%%%%%%%%%%%%%%%
%COVER PAGE
%%%%%%%%%%%%%%%%%%%%%%%%%%%%%%%%%%%%%%%%%%%%%%%%%%
\title{Astro2020 Science White Paper: The unique potential of extreme mass-ratio inspirals for gravitational-wave astronomy\vspace{-0.2cm}}

\noindent
\begin{flushleft}
%\begin{center}
{\bf Main Thematic Area:} Formation and Evolution of Compact Objects\\
{\bf Secondary Thematic Areas:} Cosmology and Fundamental Physics, Galaxy Evolution
\vspace{-0.4cm}
%\end{center}
\end{flushleft}

\begin{abstract}
The inspiral of a stellar-mass compact object into a massive ($\sim 10^{4}$--$10^{7} M_{\odot}$) black hole produces an intricate gravitational-wave signal. Due to the extreme-mass ratios involved, these systems complete $\sim 10^{4}$--$10^{5}$ orbits, most of them in the strong-field region of the massive black hole, emitting in the frequency range $\sim10^{-4}-1~\mathrm{Hz}$.  This makes them prime sources for the space-based observatory LISA (\textit{Laser Interferometer Space Antenna}). LISA observations will enable high-precision measurements of the physical characteristics of these extreme-mass-ratio inspirals (EMRIs): redshifted masses, massive black hole spin and orbital eccentricity can be determined with fractional errors $\sim 10^{-4}$--$10^{-6}$, the luminosity distance with better than $\sim 10\%$ precision, and the sky localization to within a few square degrees. EMRIs will provide valuable information about stellar dynamics in galactic nuclei, as well as precise data about massive black hole populations, including the distribution of masses and spins. They will enable percent-level measurements of the multipolar structure of massive black holes, precisely testing the strong-gravity properties of their spacetimes. EMRIs may also provide cosmographical data regarding the expansion of the Universe if inferred source locations can be correlated with galaxy catalogs. 
\end{abstract}
%
% 04.25.Nx 	Post-Newtonian approximation; perturbation theory; related approximations
% 04.30.-w 	Gravitational waves
% 04.70.-s 	Physics of black holes
% 98.62.Js 	Galactic nuclei (including black holes), circumnuclear matter, and bulges 
%\pacs{04.25.Nx, 04.30.--w, 04.70.--s, 98.62.Js}
%

\date{\today}

%%%%% AUTHOR LIST
\author{Christopher~P.~L.~Berry}
\email[]{christopher.berry@northwestern.edu; Phone: (847) 467-5076}
\affiliation{CIERA, Northwestern University, 2145 Sheridan Road, Evanston, IL 60208, USA}
%\affiliation{School of Physics and Astronomy, University of Birmingham, Edgbaston, Birmingham B15 2TT, UK}

\author{Scott~A.~Hughes}
%\email[]{sahughes@mit.edu}
\affiliation{Department of Physics and MIT Kavli Institute, Massachusetts Institute of Technology, Cambridge, MA 02139, USA}

\author{Carlos~F.~Sopuerta}
%\email[]{sopuerta@ice.csic.es}
\affiliation{Institut de Ci\`encies de l'Espai (ICE, CSIC), Campus UAB, Carrer de Can Magrans s/n, 08193 Cerdanyola del Vall\`es, Spain}
\affiliation{Institut d'Estudis Espacials de Catalunya (IEEC),  Edifici Nexus I, Carrer del Gran Capit\`a 2-4, despatx 201, 08034 Barcelona, Spain}

\author{Alvin~J.~K.~Chua}
%\email[]{alvin.j.chua@jpl.nasa.gov}
\affiliation{Jet Propulsion Laboratory, California Institute of Technology, 4800 Oak Grove Drive, Pasadena, CA 91109, USA}

\author{Anna~Heffernan}
%\email[]{anna.heffernan@ucd.ie}
\affiliation{School of Mathematics and Statistics, University College Dublin, Belfield, Dublin 4, Ireland}

\author{Kelly Holley-Bockelmann}
%\email[]{k.holley@vanderbilt.edu}
\affiliation{Department of Physics and Astronomy, Vanderbilt University and Fisk University, Nashville, TN 37235, USA}

\author{Deyan~P.~Mihaylov}
%\email{d.mihaylov@ast.cam.ac.uk}
\affiliation{Institute of Astronomy, University of Cambridge, Madingley Road, Cambridge, CB3 0HA, UK}

\author{M.~Coleman~Miller}
%\email[]{miller@astro.umd.edu}
\affiliation{Department of Astronomy and Joint Space-Science Institute,
University of Maryland, College Park, MD 20742-2421, USA}

\author{Alberto~Sesana}
%\email[]{asesana@star.sr.bham.ac.uk}
\affiliation{School of Physics \& Astronomy and Institute for Gravitational 
Wave Astronomy, University of Birmingham, Edgbaston, Birmingham B15 2TT, UK}
\affiliation{Universit\`a di Milano Bicocca, Dipartimento di Fisica G.\ Occhialini, Piazza della Scienza 3, I-20126, Milano, Italy}

\maketitle

\newpage

\setcounter{page}{1}

%%%%%%%%%%%%%%%%%%%%%%%%%%%%%%%%%%%%%%%%%%%%%%%%%%
%%%%%%%%%%%%%%%%%%%%%%%%%%%%%%%%%%%%%%%%%%%%%%%%%%
%%%%%%%%%%%%%%%%%%%%%%%%%%%%%%%%%%%%%%%%%%%%%%%%%%
\pagestyle{fancy}
\rfoot{ \thepage }
\cfoot{}
\lhead{Berry \textit{et al}.}
\rhead{The unique potential of extreme mass-ratio inspirals}

%%%%%%%%%%%%%%%%%%%%%%%%%%%%%%%%%%%%%%%%%%%%%%%%%%
%%%%%%%%%%%%%%%       MAIN TEXT        %%%%%%%%%%%
%%%%%%%%%%%%%%%%%%%%%%%%%%%%%%%%%%%%%%%%%%%%%%%%%%

\iftoggle{sections}
{\section{Low-frequency gravitational-wave astronomy}}
{\noindent{\bf I.~Low-frequency gravitational-wave astronomy.}}
\acp{GW} provide a new means to do astronomy, allowing us to observe more of our Universe and enhancing the information accessible through \ac{EM} channels. As with the \ac{EM} spectrum, different frequency bands allow us to study different systems. The first \ac{GW} observations came from the ground-based \ac{LIGO} and Virgo detectors \cite{Abbott:2016blz,LIGOScientific:2018mvr}, 
which are sensitive to the high frequency band $\sim10~\mathrm{Hz} \lesssim f \lesssim 10^3~\mathrm{Hz}$ \cite{Aasi:2013wya}. These instruments have observed the coalescences of binaries with stellar-mass ($\sim1$--$50 M_\odot$) members \cite{LIGOScientific:2018mvr,LIGOScientific:2018jsj}, and may observe the coalescence of $\sim10^2 M_\odot$ objects \cite{Abbott:2017iws}. Higher mass coalescences emit at lower frequencies. Although there are ideas for terrestrial detectors to push to lower frequencies \cite{Harms:2013raa,Abbott:2016mbw}, the best sensitivity for $3\times 10^{-5}~\mathrm{Hz}\lesssim f \lesssim 1~\mathrm{Hz}$ will come from space-borne detectors.

The \ac{LISA} will enable observation of low-frequency GWs \cite{Audley:2017drz}, bringing into view a wide range of new sources \cite[e.g.,][]{Klein:2015hvg,Caprini:2015zlo,Tamanini:2016zlh,Bartolo:2016ami,Babak:2017tow}. Among the prime anticipated targets, \ac{LISA} will observe stellar-mass systems with wide orbits (long before merger), such as the Galactic population of white dwarfs \cite{Kremer:2017xrg,Kupfer:2018jee,LittenbergWP} , and stellar-mass (and potentially intermediate-mass) binary black holes \emph{before} they become sources for ground-based detectors \cite{AmaroSeoane:2009ui,Sesana:2016ljz,CutlerWP}. 
\ac{LISA} will also detect many more massive binaries which involve \acp{MBH} of $\sim10^4$--$10^7 M_\odot$ \cite{Klein:2015hvg,Babak:2017tow,ColpiWP}.  
\acp{MBH} are thought to reside in the centers of (nearly all) galaxies \cite{Ferrarese:2004qr}. As galaxy evolution is punctuated by multiple galaxy mergers, it is expected that the \ac{MBH} within merging galaxies eventually form a binary \cite{Begelman:1980vb}. Such binaries produce extremely strong \acp{GW} which will be detectable by LISA out to redshift $z \gtrsim 20$ \cite{Klein:2015hvg}. \ac{LISA} observations of such coalescences will enable high-precision studies of the growth of \acp{MBH} across cosmic time, from their first formation at high redshift to the present day.

\begin{figure}[t!]
	\centering
	\hspace{0.4cm}
    \subfloat{\includegraphics[scale=1.0]{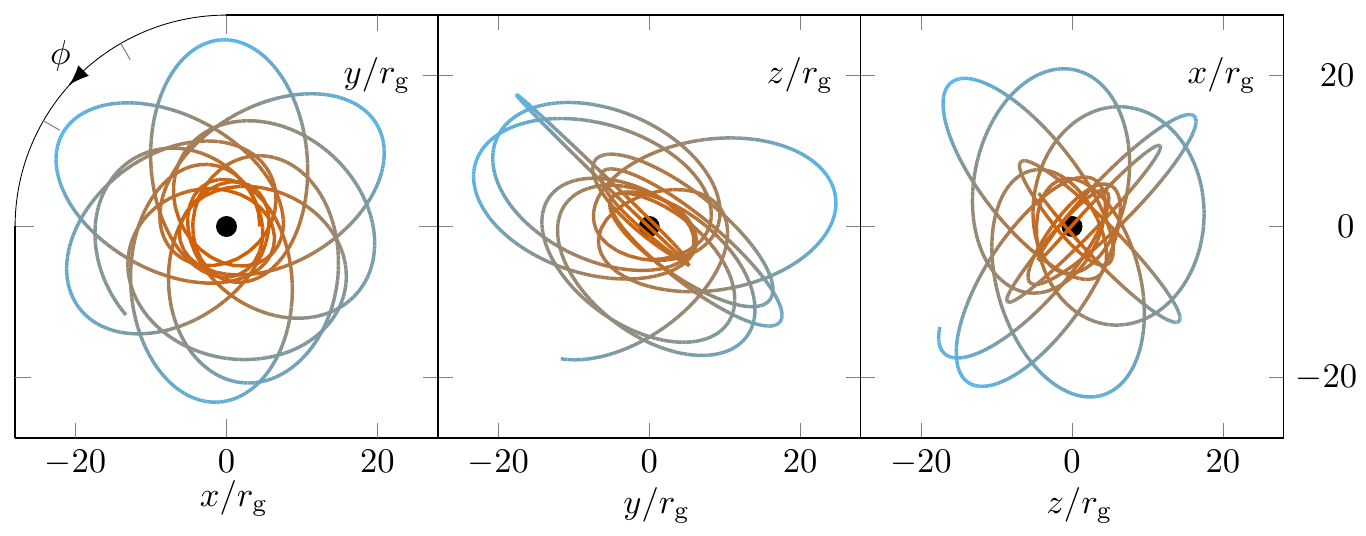}\label{fig:kerr_c}} \\
	\subfloat{\includegraphics[scale=1.0]{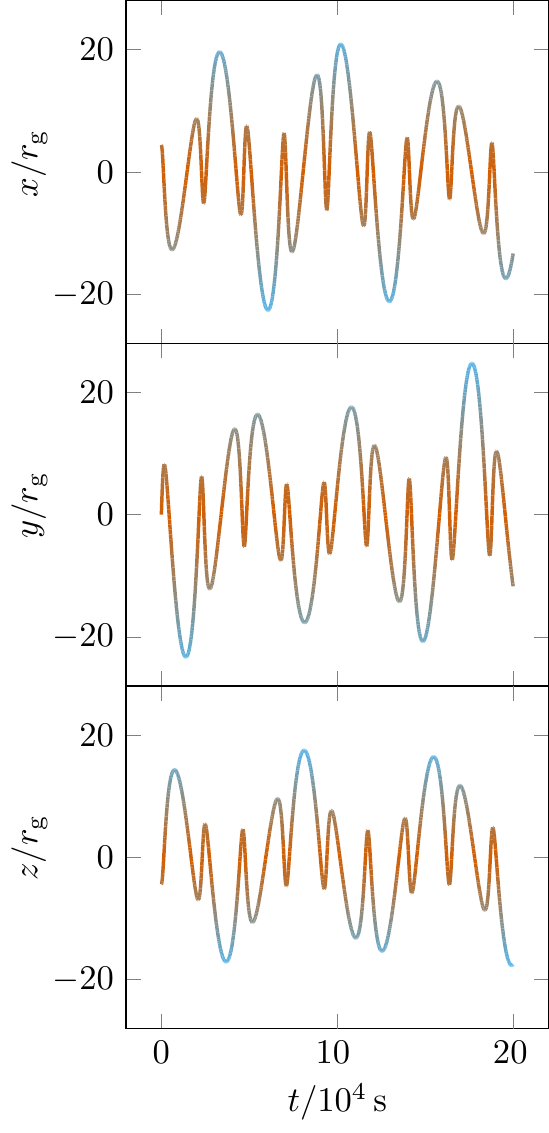}\label{fig:kerr_a}} \hspace{0.5cm}
	\subfloat{\includegraphics[scale=1.0]{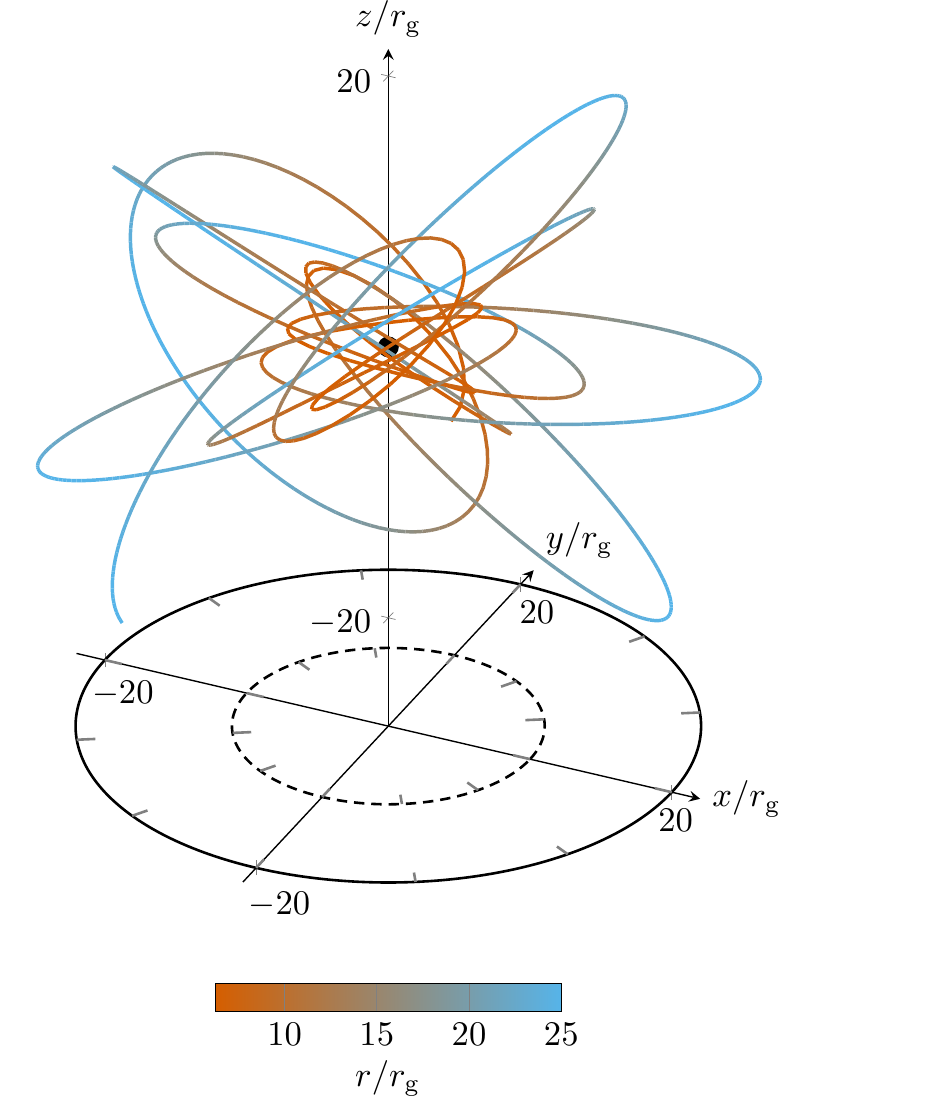}\label{fig:kerr_b}}
	\caption{Illustration of an orbit in Kerr spacetime, appropriate for a short portion of an \ac{EMRI} around a spinning \ac{MBH}. The central black hole has a mass $M = 10^{6} M_{\odot}$ and a dimensionless spin of $0.9$. Distances are measured in units of the gravitational radius $r_\mathrm{g} = GM/c^2$. The innermost stable circular orbit for this \ac{MBH} would be at $r \simeq 2.3 r_\mathrm{g}$. The coordinates have been mapped into Euclidean space to visualise the orbit: the bottom right panel shows a three-dimensional view of the orbit; the top panels show the projections of this orbit into three planes, and the bottom left panels show the orbit as a function of time. While \acp{EMRI} evolve over years, this trajectory is only a few hours long. The intricate nature of the orbit is encoded into the frequencies of the gravitational-wave signal. Measuring these lets us reconstruct the spacetime of the \ac{MBH}. Adapted from \cite{Mihaylov:2017qwn}.}
	\label{fig:kerr_geodesic}
\end{figure}

Binaries also form when stellar-mass compact objects (SCOs: white dwarfs, neutron stars or black holes) are captured by the central \ac{MBH} from the nuclear star cluster that surrounds it \cite{AmaroSeoane:2007aw}. 
These extreme mass-ratio systems will likely start on a highly eccentric orbit, emitting a burst of \acp{GW} at each pericenter passage, before evolving into a more circular inspiral during which \acp{GW} are continuously observable by \ac{LISA}. The short extreme mass-ratio bursts are only detectable from our own Galaxy and a few nearby companions \cite{Rubbo:2006vh,Berry:2013poa}, and their occurrence is expected to be rare \cite{Hopman:2006fc,Berry:2013ara}. The longer \acp{EMRI} can accumulate signal-to-noise ratio over months or years, making them detectable out to $z \sim 3$--$4$ \cite{Babak:2017tow}. \textbf{The final year before the SCO plunges into the \ac{MBH} can contain $\sim 10^{4}$--$10^5$ orbital cycles, most of which correspond to when the SCO is in the strong-field region near the \ac{MBH}'s horizon}. Tracking the evolution of the inspiral over these long durations allows for precision measurements \cite{Barack:2003fp,Babak:2017tow}. Figure~\ref{fig:kerr_geodesic} illustrates an \ac{EMRI} trajectory over a few orbits, highlighting the complicated structure of the motion. The extraordinary properties of \acp{EMRI} allow us to develop an ambitious research program that impacts astrophysics, cosmology and fundamental physics.

\iftoggle{sections}
{\section{Extreme mass-ratio inspirals and their science}}
{\vspace{2mm}
\noindent{\bf II.~Extreme mass-ratio inspirals and their science.}} 
The typical \ac{EMRI} is expected to consist of a SCO (mass $m \sim 1$--$10^2 M_{\odot}$) that inspirals into a \ac{MBH} (mass $M \sim 10^4$--$10^7 M_{\odot}$) residing within a relaxed galactic center \cite{AmaroSeoane:2007aw}. For these \acp{MBH}, the \acp{GW} generated by such \acp{EMRI} lie close to the sweet spot of \ac{LISA}'s sensitivity, near $3~\mathrm{mHz}$.

The most studied formation channel for \acp{EMRI} is the capture of a SCO from a cusp of stars onto a highly eccentric orbit \cite{Hils:1995,Sigurdsson:1996uz,Alexander:2005jz,Merritt:2006bm,Bortolas:2019sif}. \textbf{The event rates predicted for this scenario range from a low of a few events to a high of several thousand over the mission duration} \cite{AmaroSeoane:2012je,Mapelli:2012pw,Berry:2016bit,Babak:2017tow}.  
This broad range is mostly due to current uncertainties in \ac{EMRI} astrophysics, which further illustrate why there is much to learn from \ac{EMRI} measurements. Even in the most pessimistic models, \acp{EMRI} are a probable \ac{LISA} source.

\iftoggle{sections}
{\subsection{Precision astronomical probes}}
{\noindent{\bf II.A~Precision astronomical probes.}} 
Measuring the evolution of a system's orbital frequencies from the \ac{GW} signal enables us to infer, in many cases with excellent accuracy, the properties of that system.  Characteristics that affect the orbit's inspiral rate are measured with a variance that scales as $1/N_\mathrm{cyc}$, where $N_\mathrm{cyc}$ is the number of observable cycles.  For \ac{EMRI} systems, $N_\mathrm{cyc} \sim 10^4$--$10^5$, which enables exquisite measurements of system properties such as the \ac{MBH}'s mass and spin.

Babak \textit{et al}.~\cite{Babak:2017tow} studied source-parameter measurements, showing measurement precision is largely the same across different astrophysical populations. \textbf{\ac{MBH} (dimensionless) spins can be measured with a precision of $\sim10^{-6}$--$10^{-3}$, and that the redshifted \ac{MBH} masses can be measured with a fractional precision of $\sim10^{-6}$--$10^{-4}$}. Redshifted masses $M_z$ are related to physical masses $M$ by $M_z = (1+z) M$, where $z$ is the source redshift \cite{Cutler:1994ys,Holz:2005df}. To convert, we must find $z$ from the source's distance. Distance is inferred from the signal's amplitude rather than its frequency content, and so is not measured as precisely. Fractional distance uncertainties may be $\sim 0.03$--$0.3$, so source masses are measured to a similar precision \cite{Babak:2017tow}. Such mass measurements are comparable to recent results from \ac{LIGO}--Virgo \cite{TheLIGOScientific:2016wfe,LIGOScientific:2018mvr}, but for a different class of binary. The anticipated precision of spin measurements \emph{significantly surpasses} both current \ac{GW} results for stellar-mass black holes, and current X-ray measurements across a wide range of masses \cite{Reynolds:2013qqa}.

Tracking the evolution of the orbit may enable determination of the properties of the \ac{MBH}'s surroundings, showing whether the central \ac{MBH} is part of a (sub-parsec scale) binary with another \ac{MBH} \cite{Yunes:2010sm,Yang:2017aht}, or whether the compact object is embedded in a viscous environment, such as a dense accretion disk about the \ac{MBH} \cite{Levin:2006uc,Gair:2010iv,Kocsis:2011dr,Yunes:2011ws,Barausse:2014tra}.  The presence of a disk could indicate that the source is in an active galactic nucleus, which may be identifiable through \ac{EM} channels.

\textbf{Since \acp{EMRI} are long lived, the motion of the \ac{LISA} constellation allows us to localize the source, typically to within $\lesssim 10~\mathrm{deg}^2$} \cite{Babak:2017tow}. \ac{EM} follow-up may be interesting if the \ac{MBH} has an accretion disk or if the smaller component is disrupted \cite[e.g.,][]{Rathore:2004gs,Rosswog:2008ie,Sesana:2008zc,MacLeod:2014mha}. Prospects for observing \ac{EM} counterparts to white dwarfs being tidally disrupted as they inspiral into a $\lesssim 10^5 M_\odot$ \ac{MBH} are discussed in {\cite{Eracleous:2019bal}}. Multimessenger observations would give a richer insight into the properties of the system and its surroundings, as well as enabling EMRIs to be used as cosmic distance probes \cite[cf.][]{Abbott:2017xzu}. Without a counterpart, source localization will allow correlations with galaxy catalogs so that properties of the distance--redshift relation may be inferred statistically \cite{MacLeod:2007jd,CaldwellWP}.

Similarly to the \ac{MBH} mass, the redshifted compact object mass is well determined. This mass affects the rate of inspiral, and so can be determined to a precision of $\sim10^{-4}$--$10^{-7}$ \cite{Babak:2017tow}. The physical source frame masses have uncertainties dominated by uncertainty in the source redshift. As a consequence of mass segregation, we expect that \acp{EMRI} will preferentially involve the heaviest SCOs in the nuclear stellar cluster surrounding the \ac{MBH} \cite{Bachcall:1977,Alexander:2008tq,AmaroSeoane:2010bq}. However, the SCOs can span a range of masses from $\sim 0.5 M_\odot$ white dwarfs through neutron stars to stellar-mass black holes. If, in addition to black holes forming as the remnants of stellar evolution \cite{Woosley:2002zz,TheLIGOScientific:2016htt}, there are primordial black holes \cite{Carr:1974nx,Sasaki:2018dmp}, then black holes could cover the range from $\sim0.1$--$100\, M_\odot$. It may be possible to distinguish low mass primordial black holes from white dwarfs in an \ac{EMRI} by observing tidal effects during the inspiral, and perhaps a final tidal disruption \cite{Rees:1988bf,Rathore:2004gs,Ivanov:2007we,Rosswog:2008ie,Sesana:2008zc}.
If the EMRI event rate is high enough, these observations will provide a census of the SCOs residing in galactic nuclei.

\iftoggle{sections}
{\subsection{Revealing the evolution of massive black holes and their host galaxies}}
{\noindent{\bf II.B~Revealing the evolution of massive black holes and their host galaxies.}}
By precisely measuring masses and spins, \ac{EMRI} measurements will produce a catalog of these properties for an interesting population of \acp{MBH}.  Theoretical models argue that spins evolve in a characteristic manner as the \acp{MBH} grow \cite{Dubois:2013rha,Sesana:2014bea}: accretion from a disk results in high spins \cite{Volonteri:2004cf,Dotti:2012qw}, major mergers between \acp{MBH} lead to $a \approx 0.7$ \cite{Jimenez-Forteza:2016oae,Hofmann:2016yih}, and chaotic accretion of randomly orientated gas clouds, stars, and smaller black holes results in low spins \cite{Hughes:2002ei,King:2008au,Gammie:2003qi}. The catalog of \ac{EMRI}-determined masses and spins will make it possible to infer the mechanism by which these \acp{MBH} grew \cite{Berti:2008af}.

The population of \acp{MBH} in the mass range accessible to \ac{LISA} is currently not well understood \cite{Shankar:2013dwa,Saglia:2016,Savorgnan:2016,Shankar:2019yyr}. \ac{EMRI} observations will help to complete our understanding of the \ac{MBH} spectrum. \textbf{With $\sim 10$ detections, we will be able to constrain the slope of their \ac{MBH} mass function (convolved with the number of detectable \acp{EMRI} per \ac{MBH}) to a precision of $\pm0.3$} \cite{Gair:2010yu}; this precision will improve with the square-root of the number of observations. Combining insights from EMRI observations with \ac{LISA} observations of \ac{MBH} mergers \cite{Plowman:2010fc,Sesana:2010wy,Klein:2015hvg} will help guide our understanding of \ac{MBH} populations.

Since the number of \acp{EMRI} \ac{LISA} can detect depends upon the population of \acp{MBH} and the properties of their surrounding stellar clusters, the measured \ac{EMRI} rate can provide insight into the properties and dynamics of these dense systems across a range of masses \cite{Babak:2017tow}. For example, the \ac{EMRI} rate depends upon how the surrounding cluster scales with \ac{MBH} mass \cite{Mapelli:2012pw,Babak:2017tow}. \ac{EMRI} production also relies upon there being a dense cusp of stellar-mass compact objects about the \ac{MBH}. These cusps are destroyed by mergers and take time to regrow \cite{Milosavljevic:2001vi,Antonini:2015sza}. \acp{EMRI} can thus trace out how cusps regrow following disruption \cite{Babak:2017tow}.

In addition to forming by SCOs from a surrounding cusp being scattered onto highly eccentric orbits \cite{Sigurdsson:1996uz,Alexander:2005jz,Berry:2013ara}, \acp{EMRI} can form through SCO receiving fortuitous supernova kicks \cite{Bortolas:2019sif}, from the tidal break-up of binaries \cite{Miller:2005rm,Perets:2006bz,Hopman:2009gz,Antonini:2012ad,Chen:2018axp}, or from stars formed in a disk surrounding the \ac{MBH} \cite{Levin:2006uc,Stone:2016wzz}. Each channel leads to a different orbital distribution. By detecting a population of \acp{EMRI}, we will uncover the conditions governing the population dynamics of stars in the hearts of galaxies.

\iftoggle{sections}
{\subsection{Mapping black hole spacetimes}}
{\noindent{\bf II.C~Mapping black hole spacetimes.}}
The detailed information we get from \acp{EMRI} comes from the highly relativistic motion of the SCO in the strong gravitational field of the \ac{MBH}. Consequently, one of the most exciting applications for \acp{EMRI} is providing a high-precision verification of whether the \ac{GW} signal matches our expectations for a black hole system in general relativity \cite{Berti:2019xgr}. Any deviation would be evidence for new physics \cite{Gair:2012nm,Yagi:2016jml}: either these systems are not as astrophysically clean as anticipated, or our understanding of strong-field gravity is not complete. \acp{EMRI} are a unique laboratory for testing the highly relativistic, strong-field dynamics surrounding \acp{MBH}.

In general relativity, all black hole properties are fixed by its mass and spin. We can describe the background spacetime of the \ac{MBH} using a multipole expansion \cite{Geroch:1970cd,Ryan:1995wh,Li:2007qu}. The first moment is its mass; the second is its spin. The third moment (the mass quadrupole) is set by the first two \cite{Hansen:1974zz}. Measuring this moment thus allows the consistency of the Kerr solution to be checked.\footnote{For this test, all factors of the source redshift cancel, so there is no additional uncertainty from the distance measurement as there is for the source masses.} \acp{EMRI} observations enable this to be done with high precision \cite{Ryan:1997hg,Barack:2006pq}. The multipole structure may be different in alternative theories of gravity; e.g., the fourth moment differs from the Kerr prediction in dynamical Chern--Simons gravity \cite{Sopuerta:2009iy}. Bumpy black hole metrics include additional multipole structure \cite{Collins:2004ex,Gair:2007kr,Vigeland:2011ji}, and provide a theory-agnostic framework to test which deviations could be observable \cite{Glampedakis:2005cf,Chua:2018yng}. The lowest-order multipoles are best determined; \textbf{for the quadrupole moment, fractional deviations of $\sim10^{-3}$--$10^{-5}$ are discernible} \cite{Babak:2017tow}.

In addition to looking for generic deviations from the predictions for black holes in general relativity, we can consider how the \ac{EMRI} signal is affected by specific alternative scenarios. Fitting a signal to a general relativistic template will sharply constrain alternative theories, and will provide interesting places in theory space for modelers to investigate deviations from the standard picture. Specific examples that have been considered include (i) modified theories of gravity, such as Brans--Dicke gravity \cite{Berti:2005qd,Yagi:2009zm,Yunes:2011aa}, scalar Gauss--Bonnet gravity \cite{Yagi:2012gp}, $f(R)$-gravity \cite{Berry:2011pb} or dynamical Chern--Simons gravity \cite{Pani:2011xj,Canizares:2012is}, and (ii) inspirals in general relativity into an object other than a \ac{MBH} such as a massive boson star \cite{Kesden:2004qx,Macedo:2013jja} or a gravastar \cite{Pani:2010em}. The detailed information encoded within the \ac{EMRI} signal allows for precision tests of such scenarios.  For example, \ac{EMRI} constraints on dynamical Chern--Simons gravity would be four orders of magnitude better than current Solar System constraints \cite{Canizares:2012is}.

\iftoggle{whitepaper}
{}
{
\iftoggle{sections}
{\section{Facilitating the science}}
{\vspace{2mm}
\noindent{\bf III.~Facilitating the science.}}
Realizing this science requires an instrument capable of measuring \ac{EMRI} \acp{GW}, a population to observe, and the ability to find the signal in noise and to analyze its properties.  Finding and analyzing the signal in turn requires detailed models of \ac{EMRI} waveforms.

There is only one planned facility capable of observing \acp{EMRI}: \ac{LISA}~\cite{Seoane:2013qna,Audley:2017drz}. The \ac{GW} frequency range of interest is inaccessible to ground-based observatories, but \acp{EMRI} with $\sim10^4$--$10^7 M_\odot$ \acp{MBH} fall within the most sensitive portion of the \ac{LISA} band.  We now discuss in more detail the other elements needed to achieve the promise of \ac{EMRI} science.

\iftoggle{sections}
{\subsection{The number of inspirals}}
{\noindent{\bf III.A~The number of inspirals.}} 
The rate of detectable \acp{EMRI} is currently uncertain. This means that there is much to learn from EMRI measurements, but it is difficult to make predictions for how many \ac{LISA} will find. 
Building an \ac{EMRI} rate requires several astrophysical inputs:
\begin{itemize}
	\item \emph{The population of massive black holes}---Knowledge of both \ac{MBH} masses and spins are needed to predict \ac{EMRI} rates. Simulations provide insight into the population's characteristics.
For example, Babak \textit{et al}.~\cite{Babak:2017tow} used two \acp{MBH} mass distributions bracketing current observation uncertainties, one using the semi-analytic model of \ac{MBH} formation developed in \cite{Barausse:2012fy,Antonini:2015cqa,Antonini:2015sza} and one more pessimistic \cite{Gair:2010yu}. 
\ac{MBH} populations with higher spins produce more \acp{EMRI}, as do mass distributions with more \acp{MBH} in the \ac{LISA} range.
\item \emph{The distribution of stellar clusters around massive black holes}---The other component of an \ac{EMRI} is the SCO drawn from the surrounding nuclear star cluster. \ac{MBH} masses $M$ correlate with the properties of their clusters, such as the velocity dispersion $\sigma$ \cite{Ferrarese:2000se,Gebhardt:2000fk}, cluster mass \cite{Kormendy:1995er,Magorrian:1997hw}. The surrounding cluster should form a cusp with the density increasing towards the \ac{MBH} \cite{Bahcall:1976aa,Preto:2009kd}, which is ideal for \ac{EMRI} formation.  However, this cusp is disrupted by galaxy mergers, and takes time to reform \cite{Milosavljevic:2001vi,Antonini:2015sza}. In addition, the cusp and the \ac{MBH}'s properties evolve as the \ac{MBH} is fed from the cusp, changing the \ac{MBH} mass and spin, and depleting the cusp's SCO population. 
\item \emph{The range of \ac{EMRI} orbits}---An EMRI's orbit parameters strongly affect its GW signal, as well as its properties at plunge, which depend strongly upon the \ac{MBH} spin \cite{AmaroSeoane:2012cr}. These parameters must be modeled in order to estimate the expected properties of \acp{EMRI}.
\end{itemize}
Combining these ingredients with sufficient flexibility to account for current uncertainties yields Babak \textit{et al}.'s estimate~\cite{Babak:2017tow} that the intrinsic rate of \acp{EMRI} lies between a few events and several \emph{thousand} events per year. This includes EMRIs out to a redshift $z = 4.5$, which is approximately the detection horizon with a $10 M_\odot$ SCO.

\iftoggle{sections}
{\subsection{EMRI wave models and measurement analyses}}
{\noindent{\bf III.B~EMRI wave models and measurement analyses.}} 
From the modeling standpoint, the extreme mass ratio of \ac{EMRI} systems is a blessing: the spacetime is close to an exact black hole solution, and so can be analyzed using perturbation theory, using the system's mass ratio as an expansion parameter. The mass ratio also means that the inspiral is slow.  A typical \ac{EMRI} will spend tens of thousands of orbits in the \ac{LISA} band. \ac{LISA} will be able to follow the progress of each inspiral, building a detailed map of the spacetime and making highly precise measurements of the source properties \cite{Barack:2003fp,Babak:2017tow}.

Making such measurements requires waveform models which can accurately match astrophysical signals over the tens of thousands of wave cycles that will be in \ac{LISA}'s band. The \emph{self-force} program aims to develop waveforms whose systematic modeling errors are smaller than the expected statistical error associated with detector noise. This is done by calculating the SCO's gravitational field as a perturbation to the background \ac{MBH} spacetime, in the form of the effective force it produces on a geodesic orbit around the \ac{MBH} \cite{Poisson:2011nh,Blanchet:2018hut}. 
To meet the required accuracy for LISA parameter estimation, contributions to the self force at second order in the mass ratio must be computed.
Once the self force has been calculated, the orbit can be evolved, and the waveform calculated. 
Self-force calculations have advanced tremendously recently \cite{Barack:2018yvs,Barack:2018yly}, but they cannot yet produce waveform models for the most relevant cases (inclined and eccentric orbits about spinning \acp{MBH}). 
Orbit evolutions with first-order self force have been extensively studied in Schwarzschild spacetime  \cite{Warburton:2011fk,Diener:2011cc,Wardell:2014kea,Warburton:2017sxk,Heffernan:2017cad}, with the present state-of-the-art being a model for eccentric Schwarzschild orbits \cite{vandeMeent:2018rms}. 
The self force has been calculated to first order for an \ac{EMRI} moving along a generic orbit in Kerr spacetime \cite{vandeMeent:2017bcc}, and there has been extensive work towards a second-order calculation \cite{Galley:2011te,Pound:2014xva, Pound:2014koa,Miller:2016hjv,Pound:2017psq,Moxon:2017ozd}. These self-force methods, with the aid of multiscale expansions \cite{Mino:2008rr,Hinderer:2008dm,Pound:2015wva}, should ultimately produce the necessary orbit evolutions and \ac{EMRI} waveforms.

In lieu of self-force model input, the community has developed approximations (kludges) that capture the qualitative behavior of \ac{EMRI} waves to various degrees of quantitative accuracy.  The simplest models are the analytic kludge waveforms \cite{Barack:2003fp}, which approximate the orbital trajectory as a series of Keplerian ellipses supplemented with periapse precession, Lense--Thirring precession, and post-Newtonian radiation reaction. Recent work has shown that one can build an analytic kludge that better matches the frequencies of black hole orbits over longer durations while maintaining low computational cost \cite{Chua:2015mua,Chua:2017ujo}. As a consequence of their low computational costs, most analyses of how well \ac{LISA} can measure \ac{EMRI} parameters have used analytic kludges \cite[e.g.,][]{Babak:2017tow}. Another model, the numerical kludge, treats the small body's motion as an exact black hole orbit, but evolves through these orbits using approximations to the back-reaction of \ac{GW} emission \cite{Gair:2005ih,Babak:2006uv}. Although this kludge has higher fidelity than the analytic kludge \cite{Babak:2006uv,Berry:2012im}, it is significantly slower. Continuing to develop computationally efficient yet accurate waveforms remains a high priority, as the community prepares for real \ac{LISA} data.

A complication in waveform modeling is the presence of transient resonances, where the polar and radial orbital frequencies become commensurate \cite{Flanagan:2010cd}. On resonance, the usual adiabatic approximation used to calculate the orbital evolution breaks down \cite[cf.][]{Isoyama:2018sib}, and waveforms calculated this way become inaccurate. The impact of a resonance depends upon the precise value of the relative orbital phases on resonance \cite{Flanagan:2012kg,Berry:2016bit}, making it difficult to predict. Transient resonances are encountered by most inspirals \cite{Ruangsri:2013hra}; however, most are encountered late in the inspiral and when the eccentricity is low, such that failing to account for them only leads to a $\sim4\%$ loss of detectable signals \cite{Berry:2016bit}.

Although transient resonances complicate waveform modeling, they also provide an opportunity to test our understanding of gravity. Since we are sensitive to the full self-force on resonance \cite{Flanagan:2010cd,Berry:2016bit}, measuring the orbital evolution across a resonance provides a check of predictions. Furthermore, if we do not have a Kerr \ac{MBH}, the evolution around resonances can be modified, leading to sustained resonances or chaotic motion \cite{LukesGerakopoulos:2010rc,Brink:2013nna}. Observation of such behavior would be a smoking gun that the background spacetime is not Kerr, and hence the source does not contain a \ac{MBH} in a vacuum as predicted by general relativity.

Waveform models sufficiently accurate to \emph{detect} \ac{EMRI} waves will not necessarily be good enough to infer source properties.  It is not enough to have templates which match the data, we must also produce a model which can extract the source properties \emph{without bias}.  It is expected that the self-force models discussed above will eventually provide such models, but this is beyond the capability of the waveform models that we can compute today.  Much theoretical effort is needed to insure that our analyses are ready when \ac{LISA} data is available and we begin to find \acp{EMRI}.
}

\iftoggle{sections}
{\section{Summarizing the opportunity}}
{\vspace{2mm}
\noindent{\bf IV.~Summarizing the opportunity.}} 
\acp{EMRI} provide a unique means to probe the conditions in galactic nuclei and to map the spacetime of black holes. Though detection rates are uncertain, models confidently predict that they would be observed by \ac{LISA}, with potentially hundreds per year. EMRI observations would provide precision data about the masses and spins of the \ac{MBH} population, illuminating how \acp{MBH} and their surrounding galaxies evolved. \acp{EMRI} act as standard sirens, enabling measurement of the Hubble constant. The detailed information encoded in EMRI \ac{GW} signals would enable stringent tests of whether the massive objects are black holes as described by general relativity.

\textbf{\ac{LISA} is both well designed to maximize the scientific return from \acp{EMRI} and the \emph{only} planned mission able to make these revolutionary observations.}

\newpage

%%%%%%%%%%%%%%%%%%%%%%%%%%%%%%%
%%%%%     BIBLIOGRAPHY    %%%%%
%%%%%%%%%%%%%%%%%%%%%%%%%%%%%%%
%\bibliography{EMRI}

\end{document}